\documentclass{elsarticle}

\usepackage[margin=2.5cm]{geometry}
\usepackage[bookmarks=false,colorlinks=true]{hyperref}
\usepackage{amsmath,amsthm,amssymb}
\usepackage{bm}
\usepackage{fancyhdr}

\newcommand{\sfrac}[2]{{\textstyle\frac{#1}{#2}}}
\newcommand{\im}{i}
\newcommand{\op}[1]{#1}
\newcommand{\ket}[1]{|#1\rangle}

\newcommand{\yn}{\mathbb{Y}}
\newcommand{\ync}{\mathbb{Y}'}
\newcommand{\bOmega}{\bm{\Omega}}
\newcommand{\half}{\sfrac{1}{2}}
\newcommand{\onlinecite}[1]{\cite{#1}}

\makeatletter
\def\ps@pprintTitle{%
 \let\@oddhead\@empty
 \let\@evenhead\@empty
 \def\@oddfoot{}%
 \let\@evenfoot\@oddfoot}
\makeatother

\begin{document}

\title{Lowest-energy states in parity-transformation eigenspaces of $SO(N)$ spin chain}
\author{Tigran Hakobyan}
\ead{tigran.hakobyan@ysu.am,hakob@yerphi.am}
\address{Yerevan State University, 1 Alex Manoogian Street, Yerevan, 0025, Armenia}
\address{Tomsk Polytechnic University, Lenin Avenue 30, 634050 Tomsk, Russia}

\begin{abstract}
We  expand the symmetry  of the open finite-size $SO(N)$ symmetric spin chain to $O(N)$.
We partition its space of states into the eigenspaces  of  the parity transformations in the flavor space,
generating the subgroup $Z_2^{\times(N-1)}$.
It is proven that the lowest-energy states  in these eigenspaces are nondegenerate
and assemble in antisymmetric tensors or pseudotensors.
At the valence-bond solid point, they constitute  the $2^{N-1}$-fold degenerate
ground state with fully broken  parity-transformation symmetry.
\end{abstract}


\maketitle

\section{Introduction}
The spin-1 Heisenberg chain, in contrast to its spin-$\half$ analogue,
is characterized by a gap and exponentially decaying correlation as was predicted by Haldane~\cite{Haldane83}.
Its low-energy behavior is perfectly modeled by
Affleck-Kennedy-Lieb-Tasaki (AKLT) chain
with the exact valence-bond solid (VBS) ground state~\cite{AKLT}.
The Haldane phase has many fascinating properties, like
a hidden string  order parameter and  edge states~\cite{nijs}.
The two spin-$\half$ degrees of freedom at the edges
are responsible for the fourfold near degeneracy of the open chain,
which becomes exact at the AKLT point~\cite{kennedy90}. Kennedy and Tasaki (KT)
explained this behavior by a spontaneous breaking of
a hidden $Z_2\times Z_2$ symmetry, formed by
$\pi$-rotations around the coordinate axis~\cite{KT}.
They define nonlocal unitary transformation
mapping the  string order to the usual ferromagnetic order.
%
%
Recently it was observed that the Haldane phase is  protected by several discrete
symmetries~\cite{wen09,oshikawa12},
including the above $Z_2\times Z_2$~\cite{oshikawa12},
and is characterized by a double degeneracy of the entanglement spectrum \cite{oshikawa10}.

Nowadays the  spin and fermion systems with higher symmetries attract much
attention   due to experiments  in ultracold  atoms~\cite{Honercamp04}
(see Ref.~\onlinecite{rey14} for the review).
Such systems are used for the classification of  symmetry protected topological
phases \cite{KQ} and study of the entanglement spectrum \cite{rao14}.
The exact VBS ground states appear for certain
$SU(N)$ symmetric spin Hamiltonians \cite{arovas91}. 
The $SO(5)$  generalization of the AKLT chain has appeared in the context
of ladder models \cite{scalapino}.
The construction has been  extended recently to the $SO(N)$ chain
by Tu, Zhang and Xiang~\cite{xiang08}.  The authors have
revealed a  hidden antiferromagnetic order,  which is characterized by
a nonlocal string order parameter.
For the finite-size open chain, the related $Z_2^{\times (N-1)}$ symmetry,
which consists of the $\pi$-rotations in the coordinate planes,
is broken completely for odd $N$  and partially for even $N$.
Nevertheless, in both cases the ground state is $2^{N-1}$-fold degenerate.
The parity effect in $N$  persists also for the translationally invariant chain.
 In the thermodynamic limit, it is in the Haldane phase for an odd case and
 has a  twofold degenerate ground state  with broken translational
 symmetry for an even case. This fact is supported also  by the $SO(N)$ extension
 \cite{orus}  of the Lieb-Schultz-Mattis  theorem \cite{LSM61}.

In this article we consider the finite $SO(N)$ bilinear-biquadratic spin chain
with open boundaries and site-dependent  couplings.
We expand the symmetry to the $O(N)$ group by the parity transformations
(the coordinate reflections) in the flavor space.
Then partition the entire space of states  into the $2^{N-1}$
eigenspaces of the reflection operators,
which generate another subgroup $Z_2^{\times (N-1)}$.
For odd $N$ these eigenspaces  are uniquely specified the quantum numbers
of  the aforementioned $\pi$-rotation symmetry.
For even $N$ the additional $Z_2$ symmetry is needed to separate
them.
For wide range of parameters
we prove that the lowest-level state (relative ground state)
in any   eigenspace is nondegenerate.
It is proven that  the relative ground states in all eigenspaces  with $k$
odd reflection quantum numbers form $k$-th order antisymmetric tensor.
Herewith,  the parity of $k$ must coincide
with the chain length parity.
Note that the $k$-th and $(N-k)$-th order antisymmetric tensors differ
by the additional sign  under improper rotations.
For example, for $k=0,N$ both are $SO(N)$ singlets while they are, respectively,
a scalar and a pseudoscalar in $O(N)$ classification. For odd $N$,
the tensor and pseudotensor lowest-energy states alternate each other
 with the $k$ growth, while for even $N$, both multiplets appear  together.
In the $SO(3)$ case, the ground state may be, at most, fourfold degenerate
as was established  earlier by Kennedy~\cite{kennedy}.
Using the Clebsch-Gordan decomposition of the $O(N)$ spinors, constructed
by Brauer and Weyl~\cite{weyl35}, the results for general couplings are
verified at the  VBS point with the exactly known ground state.
The parity-transformation
symmetry is broken completely to $2^{N-1}$ degenerate vacua,
associated with  the relative ground states
from all eigenspaces.

\section{$O(N)$ symmetry and nonpositive basis}
The Hamiltonian of the $SO(N)$ open  spin chain  with  nearest-neighboring interaction in
vector representation has the following form:
\begin{equation}
\label{H-bracket}
\mathcal{H}=\sum_{l=1}^{L-1} \sum_{ a, b=1}^N \left(J_l\, \op{T}_{l+1}^{ a b}\op{T}_l^{ b a}
\,- \,K_l\, \op{T}_{l+1}^{ b a}\op{T}_l^{ b a}\right).
\end{equation}
Here the $T$-operators are the local projectors acting on the corresponding spin state by
$$
\op{T}^{ a b}=| a\rangle \langle b|.
$$
The spin couplings depend on site and take positive values anywhere:
\begin{equation}
\label{range}
J_l > 0, \qquad K_l>0.
\end{equation}

The Hamiltonian is invariant under $SO(N)$ rotations given by
the generators
$$
\hat{L}^{ a b} =\sum_l L^{ a b}_l,
\qquad
L_l^{ a b}=\im(\op{T}_l^{ a b}- \op{T}_l^{ b a})
$$
and can be expressed  as their bilinear-biquadratic combination.  Up to a nonessential  constant,
 \begin{equation}
 \label{H-L}
\mathcal{H}=\sum_l \big(J_l\,{\bm L}_{l+1}\cdot {\bm L}_l +K'_l (\bm{L}_{l+1}\cdot \bm{L}_l)^2\big),
\end{equation}
where
$$
K'_l=\frac{1}{N-2}(J_l-K_l),
$$
and we set for the convenience
$$
 {\bm L}_i\cdot {\bm L}_j= \sum_{ a< b}L_i^{ a b} { L}_j^{ a b}.
 $$
The range of new couplings is:
\begin{equation}
 \label{range'}
K'_l<\frac{1}{N-2}J_l.
\end{equation}
It includes the integrable
translationally invariant model at $K'=\sfrac{N-4}{(N-2)^2}J$~\cite{resh}, which generalized
Babujian-Takhtajan spin-1 integrable chain~\cite{babuj}. It also contains the
model with exact VBS ground state at $K'=\sfrac{1}{N}J$,
 considered below \cite{xiang08,xiang09}.  Recently, its low-energy effective field theory
has been studied~\cite{orus}.

Note that the symmetric combinations $\sum_l(T_l^{ a b} + T_l^{ b a})$
complement the orthogonal group to the $U(N)$.
The first term in \eqref{H-bracket} just permutes the neighboring spins  and possesses the
unitary symmetry  while the second term reduces it to the orthogonal group.
In particular, it  does not preserve the total number
of each species,
$$
\hat N_ a = \sum_l T_l^{ a a},
$$
but  preserves its parity. The latter corresponds to the
 reflection or parity transformation
 of the $a$-th flavor:
\begin{equation}
\label{parity}
\hat\sigma_ a = (-1)^{ \hat N_ a}.
\end{equation}
The commutation with the $SO(N)$ generators is
\begin{equation}
\label{sigma-L}
\hat\sigma_ c  \hat{L}^{ a b}\hat\sigma_ c
=(-1)^{\delta_{ c a}+\delta_{ c b} }  \hat{L}^{ a b}.
\end{equation}
The reflections \eqref{parity} expand $SO(N)$ symmetry of
the Hamiltonian \eqref{H-bracket}  to the entire orthogonal
group $O(N)$, which includes also improper rotations.

Evidently, the total number of spins equals the chain length, $\sum_{ a}N_ a=L$.
Therefore, the quantum numbers $\sigma_a=\pm1$ of the flavor
parity operators \eqref{parity} are subjected to the rule
\begin{equation}
\label{sigma-cond}
\sigma_1\sigma_2\ldots\sigma_N= (-1)^L.
\end{equation}
This restricts the number of independent parity transformations to $N-1$,
which are reduced in this representation to the group $Z_2^{\times (N-1)}$.
One can choose, for instance, the reflection operators of the first $N-1$
flavors as a set of its generators.

Let us focus first on the chains with even number of spins.
In this case it can be described as a quotient of the
$Z_2^{\times N}$ group, formed by independent reflections, by the $Z_2$ group,
describing the simultaneous reflection of all flavors: $\hat\sigma_1\dots\hat\sigma_N$.
Moreover, in case of  $SO(2n+1)$ symmetry, it
coincides also with the group, formed by the rotations $\hat\sigma_a\hat\sigma_b$
of the planes, spanned by the $a$-th and $b$-th flavor axis,
on the angle $\varphi=\pi$.
Indeed, due to the equation \eqref{sigma-cond},
any single reflection $\hat\sigma_a$ can be expressed in terms of the $\pi$-rotations.
 In particular, for the $SO(3)$ chain, they are described by the relation
 $\hat\sigma_1=\hat\sigma_2\hat\sigma_3$
together with two others obtained by the cyclic permutation of the indexes.
For the chains with $SO(2n)$ symmetry, the relation \eqref{sigma-cond}
reduces the number of independent $\pi$-rotations by one, so
that they form now the subgroup $Z_2^{\times(N-2)}$. An
additional $Z_2$ reflection (one can choose, for instance, $\hat\sigma_1$) complements it to the entire
parity-transformation
group.
For the odd-site chains with  $SO(2n+1)$ symmetry the generators
of the parity-transformation and $\pi$-rotation groups differ by sign, like
in the $SO(3)$ case, where we have $\hat\sigma_1=-\hat\sigma_2\hat\sigma_3$, etc.

\subsection{Nonpositive basis}
The existence of basis where the off-diagonal elements of the Hamiltonian are nonpositive
has a crucial significance for the proof of the nondegeneracy
of the lowest-energy level in the entire space of states or the subspaces specified
by good quantum numbers~\cite{M55,LSM61,LM62}.
Such  basis has no minus sign problem and can be used for Monte Carlo simulations.
Various spin and fermion lattice systems possess such basis~\cite{munro,Amb92,Shen98,Nach03}.
Usually frustration, higher-rank symmetry, or higher-order terms in the Hamiltonian
create an obstacle towards it.
Fortunately, it is possible to overcome them  for some frustrated
spin ladder systems~\cite{nakamura,H07},
as well as in one-dimension for several spin systems with higher symmetries \cite{AL86,Li01,H04,H10,harada14}.
The  $SU(N)$ and $SO(N)$ open spin chains in defining representation
have the same nonpositive basis.
It  is obtained by equipping the standard basis, composed from the single spins,
with the sign factor
\begin{equation}
\label{sign}
\theta_{ a_1\dots  a_L} = (-1)^{ \{ \#(i<j)  |  a_i> a_j \} },
\end{equation}
which counts  the number of all inversely ordered pairs of flavors and
returns its parity~\cite{H04,harada14}:
\begin{equation}
\label{basis}
\overline{| a_1\dots  a_L\rangle}  =   \theta_{ a_1\dots  a_L}   | a_1\dots  a_L\rangle.
\end{equation}
It was defined in terms of fermions by Affleck and Lieb and used
the proof of uniqueness of the ground state of $SU(2N)$ spin chain \cite{AL86}.
Later it was applied for fermion chains with various symmetries
\cite{Amb92,sorella96,LTML04,H10}.
The above form of the basis  has been applied for the extension of Lieb-Mattis theorem
 for the $SU(N)$ spin chain~\cite{H04}.
Recently, it 
 has been used in Monte Carlo simulation of $SU(N)$
\cite{mila12} and $SO(N)$ \cite{harada14} spin chains.

The sign $\theta_{ a_1\dots  a_L}$ alters under the permutation of two distinct
neighboring flavors  since they are the only pair which changes the order:
$\theta_{\dots ab\dots}=- \theta_{\dots ba\dots}$~\cite{H04}.
From the other side, it remains unchanged if two adjacent equal
spins  are replaced by another pair,
since the inclusion of double flavors
changes the amount of disordered pairs on even number:
$\theta_{\dots aa\dots}=\theta_{\dots bb\dots}$~\cite{harada14}.
Therefore, the off-diagonal matrix elements of the Hamiltonian \eqref{H-bracket},
\eqref{range} are nonpositive in the basis \eqref{basis}.

\section{Lowest-level states in $\sigma$-subspaces}
We partition the entire space of states $\mathcal{V}^L$ into $2^{N-1}$ subspaces,
 characterized by the distinct sets of the
reflection quantum numbers \eqref{parity} constrained by \eqref{sigma-cond}:
\begin{equation}
\label{V}
V^L_{\sigma_1\dots\sigma_{N}}=\{ \psi\,|\,\hat\sigma_a\psi=\sigma_a\psi\}.
\end{equation}
We call them $\sigma$-subspaces, following a similar definition
for $S^z=M$ eigen\-spaces \cite{LM62}.
The  Hamiltonian \eqref{H-bracket} remains invariant in any $\sigma$-subspace,
its  matrix is connected there in the basis \eqref{basis}.

In order to verify the last claim,  denote by $N_\pm$ the amount of the
plus and minus indexes the subspace \eqref{V}, so that $N_++N_-=N$.
According to the restriction \eqref{sigma-cond},
\begin{equation}
\label{cond}
(-1)^{N_-}=(-1)^{L}.
\end{equation}
Arrange all flavors with odd $N_a$ in ascending order:
\begin{equation}
\label{alpha-}
 a^-_1<\dots< a^-_{N_-},
\qquad \sigma_{ a_i^-}=-1.
\end{equation}
Then the Hamiltonian connects any basic state of
$V^L_{\sigma_1\dots\sigma_{N}}$
to the state
$$
\ket{\underbrace{1\dots1}_{L-N_-}a^-_1\dots a^-_{N_-}}.
$$
Indeed, acting by the first term in \eqref{H-bracket}, one can rearrange the flavors
in non-descending order. Then using the second term, one can replace
any adjacent pair $aa$ with the pair $11$. The second rearrangement gives the
desired state.

According to the Perron-Frobenius theorem \cite{Lancaster},  the lowest-energy state in
the subspace \eqref{V} is nondegenerate
and  a positive superposition of the basic states \eqref{basis} from the
subspace \eqref{V}:
\begin{equation}
\label{gs}
\Omega_{\sigma_1\dots\sigma_{N}}
=\sum_{(-1)^{N_ {a_i}}=\sigma_ {a_i}}  \omega_{ a_1\dots a_L} \,\overline{| a_1\dots  a_L\rangle}
\end{equation}
with
$$
\omega_{ a_1\dots a_L} >0.
$$
In order to detect the multiplet containing the relative ground state,
we chose a trial state.
Fill the first $N_-$ sites  by the antisymmetric combination of  the
flavors with odd numbers \eqref{alpha-}. The number of
 remaining  sites of the chain  is  even due to \eqref{cond}.
 Divide them into neighboring pairs,
 fill each pair with the same flavor, then take the sum over  all flavors. As a result,
 we arrive at the state
 \begin{multline}
\label{trial}
\Psi
=\sum_{s\in {\cal S}_{N_-}} \epsilon_{s_1\dots s_{N_-}} \, | a^-_{s_1}\dots  a^-_{s_{N_-}}\rangle
\otimes \underbrace{\psi\otimes\dots\otimes\psi}_{k}
=\sum_{s\in {\cal S}_{N_-}}\sum_{ b_1\dots b_k}  \overline{| a^-_{s_1}\dots  a^-_{s_{N_-}}
 b_1 b_1 b_2 b_2\ldots b_k b_k\rangle }
\end{multline}
with
$$
\psi=\sum_{ b}  | b b\rangle
\qquad \text{and} \qquad
k=\frac12(L-N_-).
$$
The first sum performs the antisymmetrization over all flavors with $\sigma_a=-1$
using the Levi-Civita symbol.  The constructed state is a positive superposition
of certain states from \eqref{basis},
because the insertion of a neighboring particle pair with same flavor does not change
the sign factor in \eqref{basis}.
Therefore, it has a nonzero overlap with the relative
 ground state \eqref{gs}.
Since the $\psi$ state is a scalar,
the trial state belongs
to $(N_-)$-th order antisymmetric multiplet (tensor),
which is described by the one-column Young tableau
of the same length \cite{Hamermesh},
\begin{equation}
\label{Y}
\yn_{N_-}=\yn[\underbrace{1,1,\dots,1}_{N_-}].
\end{equation}
Remember now that the nonequivalent multiplets are mutually orthogonal.
The  nondegeneracy of the relative ground state implies that it must
belong to a certain multiplet. The latter must be characterized by the same Young tableau
\eqref{Y}  due to the orthogonality condition of
nonequivalent multiplets.

In contrast to the Hamiltonian, the orthogonal symmetry mixes different $\sigma$-subspaces.
Consider the symmetric group
of permutations between different flavors, $\mathcal{S}_N\subset O(N)$.
It permutes the reflection operators  and the indexes of the  $\sigma$-subspace:
$$
s\hat\sigma_ a s^{-1}=\hat\sigma_{s( a)},
\qquad
sV^L_{\sigma_1\dots\sigma_{N}}  = V^L_{\sigma_{s(1)}\dots \sigma_{s(N)}},
$$
where $s\in \mathcal{S}_N$.
Due to   the symmetry, the Hamiltonian has the same spectrum
on all  $\sigma$-subspaces, which have the same
number of negative indexes.
We unify them into the $\binom{N}{N_-}$-fold degenerate subspace
\begin{equation}
\label{W}
\mathcal{V}^L_{N_-} \;= \;V^L_{\underbrace{-\dots-}_{N_-}\underbrace{+\dots+}_{N_+}}
\;\oplus\;
 V^L_{\underbrace{-\dots-}_{N_--1}+-\underbrace{+\dots+}_{N_+-1}}
 \;\oplus\;
  \ldots\ldots\; \oplus\;
V^L_{\underbrace{+\dots+}_{N_+}\underbrace{-\dots-}_{N_-}}\,.
\end{equation}

Thus,
\emph{the relative ground state  in the subspace $\mathcal{V}^L_{N_-}$ is a unique
 $(N_-)$-th order antisymmetric $O(N)$ tensor.
  It gathers the relative ground states  of all $\sigma$-subspaces from
\eqref{W}.
}

The entire space of states represents the sum of these degenerate subspaces:
\begin{equation}
\label{V-dec}
\mathcal{V}^L=
\begin{cases}
\mathcal{V}^L_N\oplus \mathcal{V}^L_{N-2}\oplus \ldots\oplus \mathcal{V}^L_{0}  & \text{even $N,L$},
\\
\mathcal{V}^L_N\oplus \mathcal{V}^L_{N-2}\oplus \ldots\oplus \mathcal{V}^L_{1}  & \text{odd $N,L$},
\\
\mathcal{V}^L_{N-1}\oplus \mathcal{V}^L_{N-3}\oplus \ldots\oplus \mathcal{V}^L_{1} & \text{even $N$, odd $L$},
\\
\mathcal{V}^L_{N-1}\oplus \mathcal{V}^L_{N-3}\oplus \ldots\oplus \mathcal{V}^L_{0} & \text{odd $N$, even $L$}.
\end{cases}
\end{equation}
It must be mentioned that the subspace
$\mathcal{V}^L_{N_-}$  is not empty if $N_-\le L$. So we suppose in the following that the chain
length is large enough, $L\ge N$.

\emph{The total ground state  may be, at most, $2^{N-1}$-fold degenerate
combining the lowest-level multiplets from all subspaces $\mathcal{V}^L_k$ in the decomposition \eqref{V-dec}.}

According to the representation theory of orthogonal
algebras  \cite{Hamermesh},  two multiplets, described by the Young diagrams
$\yn_{N_\pm}$, are mutually conjugate and related by the  Levi-Civita symbol
$\epsilon_{ a_1\dots a_N}$.
The conjugate multiplets are distinguished by the sign under improper rotations,
which maps tensor to  pseudotensor.
For example,  $\yn_0$ is a scalar (singlet) while $\yn_N\sim \ync_0$
is a pseudoscalar;  $\yn_1$ is a vector while $\yn_{N-1}\sim \ync_1$
is a pseudovector, etc.
In particular, the lowest level state is a scalar
 in the subspace $\mathcal{V}^L_0$ and a  pseudoscalar  in $\mathcal{V}^L_{N}$.
As $SO(N)$ representations, they  are  equivalent,
and the smallest number  from the set $N_\pm$  characterizes the multiplet.
According to \eqref{V-dec}, the distribution on $k$ of the lowest-level multiplets in
$\mathcal{V}^L_k$ depends sharply on the parity of $N$.
In an odd case, the tensors and pseudotensors alternate each other
 with the growth of $k$, while in an even case, both type multiplets
 appear  together. The self-conjugate representation $\yn_{N/2}$
 emerges only once for $N$ being a multiple of $4$.

Now turn back to the  another $Z_2^{\times (N-1)}$ symmetry,
formed by the $\pi$-rotations in the $\binom{N}{2}$ coordinate planes,
$$
\hat\sigma_ a\hat\sigma_{ b}=e^{i\pi \hat{L}^{ a b}}.
$$
The $\sigma$-subspaces are the  eigenspaces of its elements
corresponding to the quantum numbers $\sigma_a\sigma_b$. They
remain unchanged under the simultaneous change of all signs
$\sigma_ a\to -\sigma_ a$ and, therefore, do not distinguish
between  conjugate representations $\yn_{N_\pm}$. For example, both the scalar
and pseudoscalar are labeled by the plus signs.
For odd $N$, the elements of this group
parameterize uniquely the  $\sigma$-subspaces, since due to the condition
\eqref{cond} among the two subspaces
$V^L_{\sigma_1\dots\sigma_N}$ and  $V^L_{-\sigma_1\dots-\sigma_N}$,
only one exists for a given chain.
In contrast, for even $N$, both subspaces present or absent  simultaneously and
and belong to the subspaces $\mathcal{V}^L_{N_\mp}$ respectively.
An additional $Z_2$   quantum number $\sigma_1$ separates them.

\section{$SO(3)$ case}
Consider the simplest case of  $SO(3)$ chain and express the bilinear-biquadratic Hamiltonian
\eqref{H-L} through the standard spin-one operators:
 \begin{gather}
 \label{H-S}
\mathcal{H}=\sum_l \big(J_l\,{\bm S}_{l+1}\cdot {\bm S}_l +K'_l (\bm{S}_{l+1}\cdot \bm{S}_l)^2\big),
\end{gather}
with $S^ a_l=-\epsilon^{ a b c}L^{ b c}_l$ and $K'_l<J_l$.
 According to the general case \eqref{W},
 the $\sigma$-subspaces are unified into four degenerate subspaces,
\begin{equation}
\label{W-3}
\begin{aligned}
  \mathcal{V}^L_0 = V^L_{+++}, \qquad  \mathcal{V}^L_2 = V^L_{+--}\oplus V^L_{-+-}\oplus V^L_{--+},
\\
 \mathcal{V}^L_3=V^L_{---}, \qquad \mathcal{V}^L_1 = V^L_{++-}\oplus V^L_{+-+}\oplus V^L_{-++},
\end{aligned}
\end{equation}
so that the first or second lines happen, respectively, for the chains with even or odd lengths.
Following \eqref{V-dec}, the entire space of states can be expressed in terms of them
as follows:
\begin{equation}
\label{V-W-3}
\mathcal{V}^L=
\begin{cases}
\mathcal{V}^L_2\oplus \mathcal{V}^L_0 & \text{for even $L$},
\\
\mathcal{V}^L_3\oplus \mathcal{V}^L_1 & \text{for odd $L$}.
\end{cases}
\end{equation}

Applying the previously obtained results for $N=3$ case, we conclude that
\emph{the lowest-energy states in the subspaces $\mathcal{V}^L_0$ and $\mathcal{V}^L_3$
are spin-singlets, which behave as a scalar  and pseudo-scalar, correspondingly, under improper
rotations.
In the subspaces $\mathcal{V}^L_1$ and $\mathcal{V}^L_2$ they form  spin-triplets with vector
and pseudo-vector behavior. All these multiplets are nondegenerate.}

The total ground state is either a unique spin-singlet, or a unique
spin-triplet, or their superposition.
The first opportunity happens for  $K'_l\le 0$~\cite{munro}, and
the last one takes place at the AKLT point  $K'_l=\sfrac{1}{3}J_l$.
So, it  may be at most fourfold degenerate
as was established already by Kennedy in Ref.~\onlinecite{kennedy}.
He used another partition and another negative basis. The latter is obtained from  the
usual Ising basis by a nonlocal unitary shift locally equivalent to the KT transformation.
Its relation with the basis \eqref{basis} has been studied in detail recently~\cite{harada14}.

\section{Exact VBS case}
The  coupling values $K'_l=\sfrac{1}{N}J_l$  describe the chain with the
exact $2^{N-1}$-fold degenerate VBS ground state \cite{xiang08,xiang09}.
The couplings can be made site-dependent  since the ground state minimizes apart all local interactions.
The spinor representation of the orthogonal group has been used in order to obtain the explicit
expression of the ground-state \cite{scalapino,xiang08,xiang09}
in the matrix product form~\cite{fannes}.
Here we adopt the construction to the open chain and show that the parity-transformation
symmetry $Z_2^{\times(N-1)}$ is broken completely.

Define  $2^n$-dimensional  gamma matrices with $n=[\sfrac{N}{2}]$, which
generate the Clifford algebra
$$
\{\Gamma^ a,\Gamma^ b\}=2\delta_{ a b}.
$$
The spinor representation of $SO(N)$ is given by
$$
L^{ a b}=-\frac{i}{2}[\Gamma^ a,\Gamma^ b].
$$
The rotation $R$
in the flavor space induces the unitary transformation of spinors:
\begin{equation}
\label{UR}
U_R\Gamma^ a U^+_R=\sum_ b R_{ a b}\Gamma^ b.
\end{equation}
The ground state  is presented in the matrix product form \cite{scalapino},
\begin{equation}
\label{MP}
\Omega_{\alpha\beta} \sim \sum_{ a_1, a_2,\dots, a_L}
(\Gamma^{ a_1}\Gamma^{ a_2}\dots \Gamma^{ a_L})_{\alpha\beta}
\ket{ a_1 a_2\ldots a_L},
\end{equation}
where $\alpha,\beta$  are $2^{n}$-dimensional spinor indexes.
The trace gives rise to the translationally invariant ground state \cite{xiang08}.
As a result, the ground state transforms under rotations as
$$
\bOmega\to U_R\bOmega U^+_R,
$$
and, hence, belongs to
the tensor product of the spinor representation $\Delta$ and its dual $ \Delta^* $
(which are equivalent). Its Clebsch-Gordan decomposition depends on whether  $N$ is even or odd.

For $O(2n)$, it reads~\cite{weyl35}:
\begin{equation}
\Delta\otimes\Delta^*=
\bigoplus_{k=0}^N \yn_k=\bigoplus_{k=0}^{n-1} (\yn_k\oplus\ync_k)\oplus\yn_n.
\end{equation}
The spinor representation \eqref{UR} expands to the $O(2n)$ representation by
$\hat\sigma_a=\Gamma^a\Gamma^0$. Here
$$
\Gamma^0=\imath^n\prod_{a=1}^N\Gamma^a,
$$
where $\imath$ is $1$ or $i$ according as $n$ is even or odd.
The multiplets $\yn_k$ and $\ync_k$ are
formed by the components of the rank $k$ antisymmetric tensor,
\begin{align}
\label{omega}
\Omega^{b_1\dots b_k}_{\yn_k}&=\text{Tr}(\bOmega\Gamma^{[b_1}\dots\Gamma^{b_k]}),
& k\le n,
\\
\label{omega'}
\Omega^{b_1\dots b_k}_{\yn'_k}&=\text{Tr}(\bOmega\Gamma^0\Gamma^{[b_1}\dots\Gamma^{b_k]}),
& k<n,
\end{align}
where the antisymmetrization is performed over the indexes in square brackets.
Since the trace of a product of odd number of gamma matrices vanishes,
only the states obeying \eqref{cond} survive
in the sum \eqref{MP}. This reduces the number of
nonvanishing matrix elements of $\Omega_{\alpha\beta}$ from $2^N$ to $2^{N-1}$,
as was also mentioned in
Ref.~\onlinecite{harada14}.
The trace of a product of even number of gamma matrices
is expressed in terms of the Kronecker delta products, which
vanish unless their indexes coincide pairwise.
Each index $b_i$ from the trace decomposition
in \eqref{omega} will be paired by $\delta^{b_ia_j}$ with any index $a_j$ from \eqref{MP}.
The antisymmetrization eliminates the pairings between the indexes $b_i$.
As a result, we obtain a combination of states with $k$ distinct flavors
$b_1,\dots,b_k$ and  $L-k$ flavors partitioned into the singlets $\psi$ from
\eqref{trial}.   Therefore, the ground state
$$
\Omega^{b_1\dots b_k}_{\yn_k}=\bOmega_{\yn_k}
$$
belongs to the subspace
$\mathcal{V}^L_k$ defined in \eqref{W}.
Moreover, the  $2^{N-1}$-fold degenerate ground state
splits into them in complete agreement with the decomposition \eqref{V-dec} for even $N$:
\[
 \bOmega=
\begin{cases}
\bigoplus_{i=0}^n\bOmega_{\yn_{2i}}=
 \bOmega_{\yn_0}\oplus\bOmega_{\ync_0}\oplus\bOmega_{\yn_2}\oplus\bOmega_{\ync_2}\oplus\ldots
 \\
\bigoplus_{i=1}^{n}\bOmega_{\yn_{2i-1}} =
\bOmega_{\yn_1}\oplus\bOmega_{\ync_1}\oplus\bOmega_{\yn_3}\oplus\bOmega_{\ync_3}\oplus\ldots,
\end{cases}
\]
where the first and second lines correspond, respectively, to even and odd values of $L$.
The parity-transformation symmetry is fully broken. In contrast, the $\pi$-rotation symmetry
is broken partially \cite{xiang08}, up to the mutually conjugate multiplets.
So, like for the $\sigma$-subspaces, one needs in a single $Z_2$ reflection in order to separate
the tensors from pseudotensors and get a fully broken symmetry.

For $O(2n+1)$, the Clebsch-Gordan decomposition  is ~\cite{weyl35}
\begin{equation}
\label{clebsch-odd}
\Delta\otimes\Delta^*=\bigoplus_{i=0}^n \yn_{2i}=\yn_0\oplus\ync_1\oplus\yn_2\oplus\ync_3\oplus\ldots,
\end{equation}
where the second sum concludes with $\yn_{n}$ or $\ync_{n}$, respectively, for  even or odd $n$.
Now $\Gamma^0$ commutes with $\Gamma^a$ and is a number.
A product with nonzero trace exists for any number ($\ge N$) of gamma matrices.
The simultaneous flip in the sign of all flavors
now is an improper rotation  acting by
$$
\Psi\to (-1)^L\Psi
$$
on all states \eqref{sigma-cond},
including the ground state \eqref{MP}.
This modifies the decomposition \eqref{clebsch-odd}  for odd-length chains:
\[
 \bOmega=
\begin{cases}
\bigoplus_{i=0}^n\bOmega_{\yn_{2i}}=\bOmega_{\yn_0}\oplus \bOmega_{\ync_1}\oplus \bOmega_{\yn_3}\oplus\ldots
& \text{even $L$},
 \\
\bigoplus_{i=0}^{n}\bOmega_{\ync_{2i}} =
\bOmega_{\ync_0}\oplus \bOmega_{\yn_1}\oplus \bOmega_{\ync_3}\oplus\ldots
& \text{odd $L$}.
\end{cases}
\]
This sum corresponds to the superposition of the lowest-level multiplets
inherited from the decomposition \eqref{V-dec} for odd $N$.
The  $SO(N)$ decomposition is obtained upon the substitution  $\ync_k=\yn_k$
in both sums  \cite{xiang09}.
Like the $\pi$-rotation symmetry \cite{xiang08},
the parity-transformation symmetry is broken completely.

The AKLT Hamiltonian corresponds to the $O(3)$ symmetric spin-1 chain \eqref{H-S}
at $K'_l=\sfrac{1}{3}J_l$~\cite{AKLT}.
The fourfold degenerate ground state splits into singlet and
triplet states. For the even $L$, they are lowest-energy states of $\mathcal{V}^3_0$ and  $\mathcal{V}^3_2$
and represented by a scalar and pseudovector, respectively. For the odd $L$, they are lowest-energy states of
$\mathcal{V}^3_1$ and  $\mathcal{V}^3_3$
and represented by a vector and pseudoscalar, respectively.

\section{Conclusion}

In the current article,  the properties of the lowest-energy states of the $SO(N)$
bilinear-biquadratic spin chain with $L$ spins and open boundaries are studied  for wide range of couplings.
We consider the reflections with respect to each flavor (parity transformations),
which  generate  the
$Z_2^{\times (N-1)}$ group, expanding the symmetry to $O(N)$.
It
splits    the entire space of states  into the $2^{N-1}$ subspaces $V_{\sigma_1\dots\sigma_N}$,
each characterized by the set of $N$ reflection quantum numbers $\sigma_a=\pm$,
subjected to the condition
\eqref{sigma-cond}.

For wide range of parameters it is proven that the lowest-level state
in any  such $\sigma$-subspace is nondegenerate.
Moreover,  the lowest states from all subspaces
$V_{\sigma_1\dots\sigma_N}$
with $k$ negative indexes
form $k$-th order antisymmetric tensor, or $O(N)$ multiplet.
In particular, the relative ground state is a scalar and pseudoscalar,
respectively, in the subspaces with positive ($V_{++\dots +}$)
and negative ($V_{--\dots -}$) signs.

As a result, the entire ground state may be, at most, $2^{N-1}$-fold degenerate.
The maximal degeneracy appears at the point with the exact valence-bond solid
(VBS) ground state, where the $Z_2^{\times (N-1)}$ symmetry
is broken completely to $2^{N-1}$ degenerate vacua,
described by the relative ground states from all $\sigma$-subspaces.
The even-odd effect both on the group rank and chain length is observed.
For even $N$, the degenerate ground state is formed by $k$-th order
tensor and pseudotensor multiplets.
For odd $N$, tensor and pseudotensor multiplets alternate each other:
$k$-th order tensor precedes the $(k+1)$-th order pseudotensor.
In both cases, the parity of $k$ has to coincide
with the parity of the chain length $L$.
In particular, the ground state of the usual spin-1 bilinear-biquadratic chain with
even number of spins is the superposition of a scalar and a pseudovector,
while for odd-site chain  it combines a pseudoscalar and a vector.

Out of the VBS point, the degeneracy between the different multiplets is removed.
However,  any such multiplet still combines all lowest-level states
from the $\sigma$-subspace with the same set of
reflection quantum numbers $\sigma_a$.
This property sheds light on the origin of the broken $Z^{\times (N-1)}$
symmetry at the VBS point.

\section*{Acknowledgments}
This work was partially supported by the Armenian State Committee of Science
grant No. 15RF-039  and by the Armenian National
Science \& Education Fund (ANSEF) grant No.~3501.
It was done within the Regional Training Network Program on Theoretical Physics, sponsored by
the Volkskswagen Foundation of Germany, contract No. 86~260,
and  within the International Centre of Theoretical Physics (ICTP) Network Project NET~68.

\end{document}